\documentclass{egpubl}
\usepackage{pgv17}

\WsPaper         
\electronicVersion 

\ifpdf \usepackage[pdftex]{graphicx} \pdfcompresslevel=9
\else \usepackage[dvips]{graphicx} \fi

\PrintedOrElectronic

\usepackage{t1enc,dfadobe}

\usepackage{egweblnk}
\usepackage{cite}

\usepackage{epstopdf}
\usepackage{etoolbox}
\usepackage{mathtools}
\usepackage{xcolor}
\usepackage{array, booktabs, caption}
\usepackage[flushleft]{threeparttable}
\usepackage{intcalc}
\usepackage{amsmath}
\usepackage{tikz}
\usepackage{algorithm}
\usepackage[noend]{algpseudocode}
\usepackage{subcaption}
\usepackage{soul}
\usepackage[utf8]{inputenc}
\usepackage{amsfonts}

\newcommand{\impmark}{\strut\vadjust{\domark}}
\newcommand{\domark}{%
  \vbox to 0pt{
    \kern-\dp\strutbox
    \smash{\Large\llap{*\kern1em}}
    \vss
  }%
}

\newcommand{\todo}[1]{}
\renewcommand{\todo}[1]{{\color{red} TODO: {#1} \\}}

\newcommand{\notes}[1]{}
\renewcommand{\notes}[1]{{\color{blue} Notes: {#1} \\}}

\newcommand{\highlight}[1]{}
\renewcommand{\highlight}[1]{{\impmark\color{red} {#1} }}

\def\abstract{
  \@abstract}
\def\endabstract{
  \endlist
 }
\def\@abstract{\list{}{\leftmargin 0pc\rightmargin\leftmargin
  \parsep 0pt plus .1pt}\item[]{\textbf{Abstract}}\\\itshape}
  
\usetikzlibrary{plotmarks}

\title[Photo-Guided Exploration of Volume Data Features]%
      {Photo-Guided Exploration of Volume Data Features}

\author[M. Raji, A. Hota, R. Sisneros, P. Messmer \& J. Huang]
       {Mohammad Raji$^1$, Alok Hota$^1$, Robert Sisneros$^2$, Peter Messmer$^3$ and Jian Huang$^1$
        \\
         $^1$University of Tennessee, Knoxville, Tennessee, United States\\
         $^2$National Center for Supercomputing Applications, Urbana, Illinois, United States\\
         $^3$NVIDIA, Zurich, Switzerland
       }

\begin{document}


\maketitle

\begin{abstract}
In this work, we pose the question of whether, by considering qualitative information such as a sample target image as input, one can produce a rendered image of scientific data that is similar to the target. The algorithm resulting from our research allows one to ask the question of whether features like those in the target image exists in a given dataset. In that way, our method is one of imagery query or reverse engineering, as opposed to manual parameter tweaking of the full visualization pipeline. For target images, we can use real-world photographs of physical phenomena. Our method leverages deep neural networks and evolutionary optimization. Using a trained similarity function that measures the difference between renderings of a phenomenon and real-world photographs, our method optimizes rendering parameters. We demonstrate the efficacy of our method using a superstorm simulation dataset and images found online. We also discuss a parallel implementation of our method, which was run on NCSA's Blue Waters. 



\thispagestyle{empty}
\end{abstract}

%

%

	
\section{Introduction}

As exemplified by Google Images, web-scale online image resources have become very powerful over the past decade. These online resources can help people find, share and reuse observational photographs of natural phenomena in unprecedented ways.  

Nowadays it is easy for a user to query, for example, “Hurricane Isabel” in a search engine and find satellite photographs of said hurricane. With some reasonable efforts of quality checking, the resulting images are indeed useful to users who wish to learn more about “Hurricane Isabel”.  

While this practice is already widespread, in an exciting way there is another emerging trend -- scientific simulation data are becoming more accessible as well.  Now, to a user who has access to a computationally modeled dataset, it is not rare and not unreasonable to ask, “does this dataset have visual features that match those in the photographs queried online?”

That question is not new.  The traditional approach to answer that question, however, can have a variety of barriers related to the expertise, skill, and time that are required to set up, control, and use a scientific visualization pipeline effectively. The interactive visualization process is human driven, because refining visualization parameters requires a human expert to visually examine the renderings and mentally translate the observed differences into the fine tuning of parameters in the visualization pipeline.

In this work, we pose the following question. Given a set of observational photographs as sample input, how can an algorithm drive the refinement of parameters in a visualization pipeline so that the volume rendered images in the end match the visual features in the photographs? 

\begin{figure}
	\centering
	\includegraphics[width=0.7\linewidth]{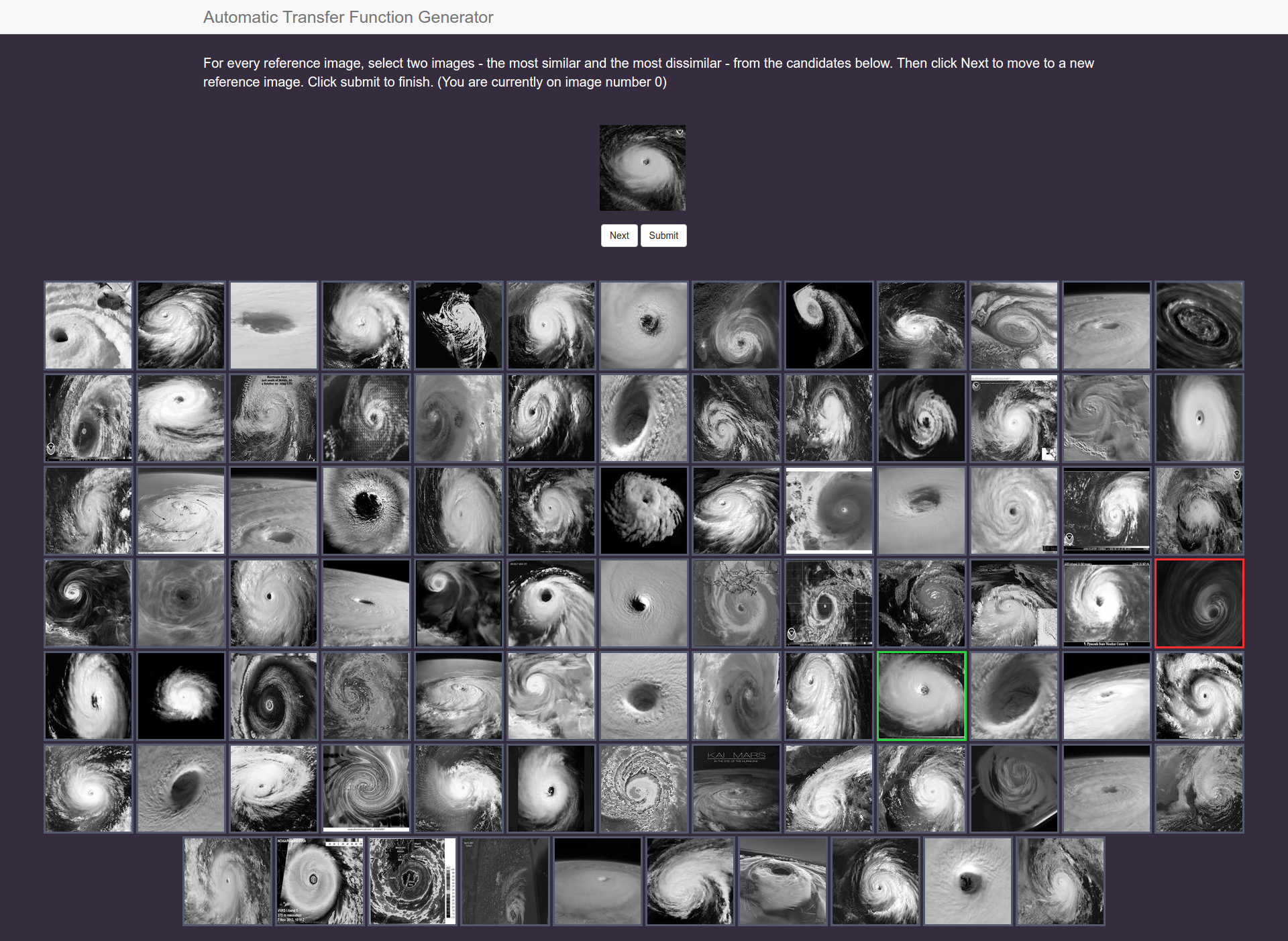}
	\caption{The web-based interface used for training the Siamese network. The interface shows a random reference image on the top of the screen. In this example, one similar and one dissimilar image have been selected by the user.}
	\label{fig:web}
\end{figure}

We set up our experiment based on standard practices in the field. The visualization package we chose is VisIt \cite{VisIt}.  We chose VisIt because of its implementation of the scientific visualization pipeline and its wide adoption among research and industry users. For sample input, we used 89 hurricane satellite images queried from Google Images (Figure \ref{fig:web}). For the computationally modeled dataset, we used a WRF superstorm simulation (48 time steps, $254 \times 254 \times 37$ spatial resolution and the humidity variable).

The best approach we have found is a combination of genetic algorithm and deep learning. While using genetic algorithms to auto-tune scientific visualization has been explored before \cite{he1996generation}, combining genetic algorithm together with deep learning is unique.  In particular, we need a robust metric to compare the similarities between volume rendered images and sample input images.  The deep learning based image comparison metric allows us to start with a population of primitive rendering parameters (i.e. transfer functions), and stochastically iterate the whole population towards the target via crossover and mutate operations.

In a repeatable manner, our experiment succeeded in creating volume rendered images that can match the visual features in the examples from Google Images. We aim to validate the hypothesis that the otherwise human-centered task of visual search can be automated. By controlling the search space and leveraging parallel computing, our experiments typically complete within 6-7 minutes using 60 compute nodes on the Blue Waters supercomputer \cite{bluewaters}. 

The VisIt package is used entirely from its Python interface. The final volume rendered images that our experiments produced are grayscale because we specifically focus on the effect of opacity on feature occlusion. The renderings are from a typical top-down viewing frustum, because satellite images have that typical view point. The deep learning component was implemented in Python using TensorFlow \cite{tensorflow2015-whitepaper}.




In the remainder of the paper, we summarize the background work in Section 2 and describe details of our approach in Section 3.  We present results in Section 4. Discussions, conclusion and potential directions for future work are in Sections 5 and 6.

\section{Background Work}

When assuming an unlimited amount of time and an unlimited amount of computing resources, the brute force way to achieve our goal is simply to vary every parameter through the parameter space exhaustively. Our work is to accelerate that process by making the search process more intelligent and more robust. From this respect, our specific aim is different from, although related to, transfer function design.

Most work on transfer function design relies on developing a deeper understanding of the characteristics in the volume data. 
%
For example, segmentation- and clustering-based methods 
\cite{maciejewski2011abstract, zhang2016clustering, sereda2006automating, selver2009semiautomatic, nguyen2011automatic, roettger2005spatialized} 
work by applying clustering algorithms on data histograms in search of defined boundaries in the dataset.
They work well for medical datasets. 
For another example, visibility-based methods consider the camera viewpoint and its effect on occlusions. Correa and Ma introduced the idea of visibility histograms as a way to show the visibility of different regions in a volume \cite{correa2009visibility, cai2013automatic}. These semi-automatic techniques usually require quantitative input from the user in order to know which parts of the volume should be exposed and to what degree. 

More related to our work are image-based methods, 
which evaluate a transfer function by evaluating the rendered images, and cast transfer function design improvement as an optimization problem~\cite{he1996generation, wang2007intelligent}. However, few previous methods approached transfer function design with a target in mind. Furthermore, few previous methods required an ability to distinguish whether a newly rendered image is incrementally better, beyond being just qualitatively different. 
To our knowledge, our work herein is the first that attempted to address the potential to procedurally guide the refinement of a visualization towards a set of provided samples.

%

	

	There are many image similarity measures in the literature of computer vision and machine learning \cite{di1999distance, bar2003learning, yang2006efficient, frome2007learning, lee2008rank, morel2009asift}. Many methods use local features in an image to create a similarity distance. For example, Morel and Yu used local SIFT features to create a distance metric \cite{morel2009asift}. These methods tend to focus on low-level features as opposed to high-level semantics, therefore working best for images that are already similar. Our need is different because we have to start with image pairs that may be very different at first, and only gradually become more similar after each iteration. We need similarity metrics that work better on a higher level.

	
	We explored deep learning, because of its proven efficacy with image classification \cite{krizhevsky2012imagenet}, the available variety of neural networks~\cite{druzhkov2016survey}, and measuring image similarity
\cite{wang2014learning, hoffer2015deep, zagoruyko2015learning, taigman2014deepface}. 
	
For our experiment, we chose a deep neural network (DNN) called a Siamese network for making image comparison in our system. This type of network was originally used for verifying signatures \cite{bromley1993signature}. Siamese networks are commonly used to measure the difference between images. They take pairs of images as input and learn to increase their output value for dissimilar images and decrease it for similar images. A typical concern in using neural networks is properly fitting the model with enough training data and not \textit{overfit} the network. However in our use case, we can tolerate overfitting. Our system does not need high accuracy, but only needs to lead the optimization in the correct direction. Additionally, our optimizer tweaks input to the neural network's benefit. Details of our Siamese network as an image comparison metric are in Section \ref{sec:metric} and a discussion on overfitting is in Section \ref{sec:discussion}.

Our work also bears similarity to \cite{rheingans2001volume} and \cite{lu2005example}, where the authors pioneered an effort to tune existing volume renderings to match appearances, as exemplified in selected medical illustrations. Their method focused on rendering style and did not involve redesigning transfer functions. Our work is solely about the search and discovery in the complex transfer function space. Our sample targets, being photographs, can also be very different from what they may be compared to. 
	
\section{Method}

Our experiment follows a two-step process. First, a user uses a web-based interface to train the Siamese network into a similarity metric. The metric is specifically customized for the kind of example images queried from online sources based on keywords (in this work: hurricane satellite image). This step requires user input and typically lasts 30 minutes. Second, the user chooses a viewpoint for all renderings. In our experiment, we chose a top down viewpoint because this is typical of the satellite images used. 


 
\begin{figure}
	\includegraphics[width=\linewidth]{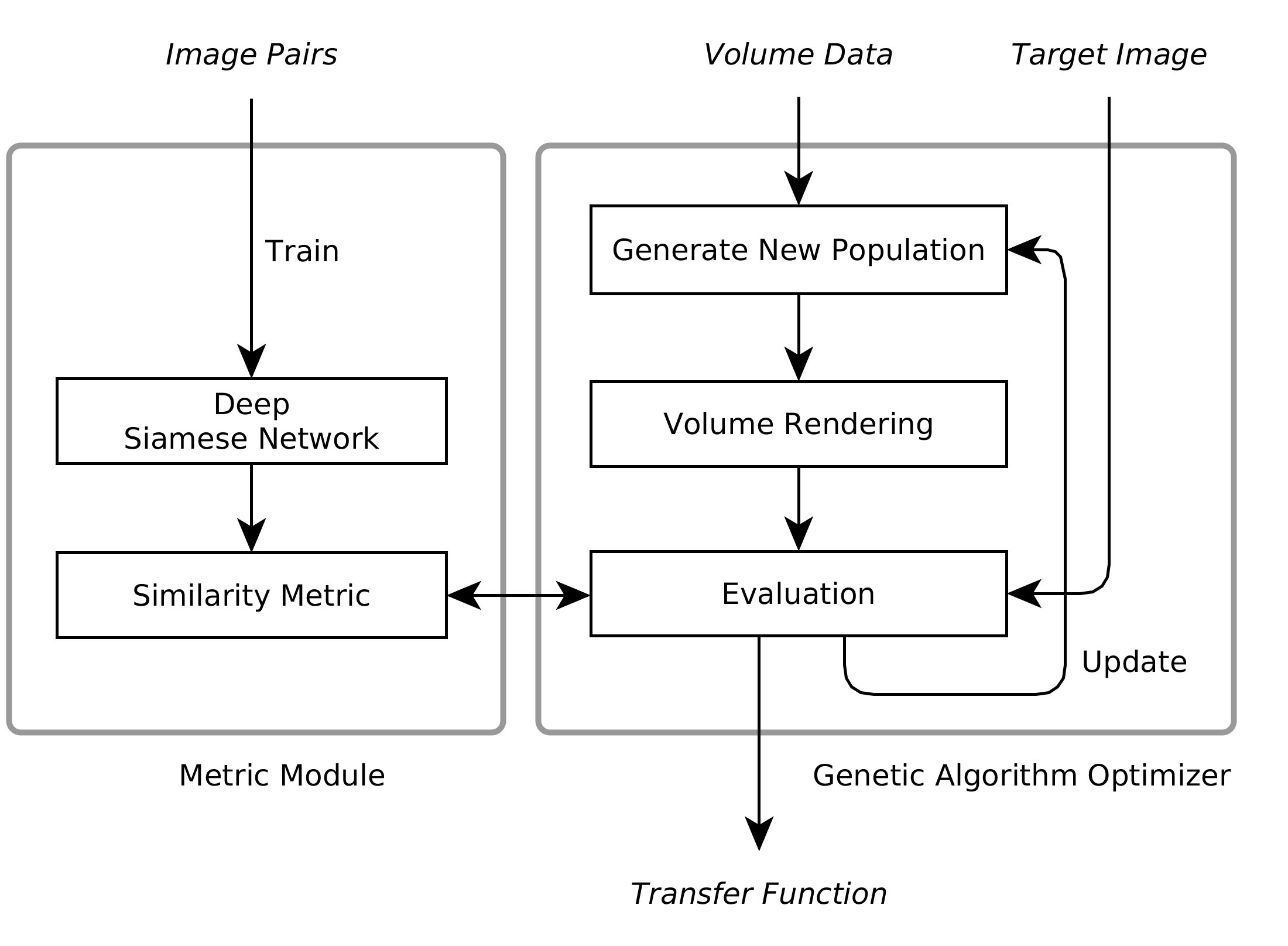}
	\caption{The main architecture of our system. A deep Siamese network is trained on pairs of images related to a phenomenon and a similarity metric is created. The similarity metric then drives the genetic algorithm optimizer to find a transfer function. }
	\label{fig:overview}
\end{figure}

The block diagram of our method is in Figure \ref{fig:overview}. There are two conceptual modules:  the \textit{metric} module on the left and the genetic algorithm \textit{optimizer} on the right. The metric module is pre-trained and then used in the optimization process for comparing images to the data renders. The optimizer traverses the search space and, using the similarity metric from the metric module, finds the optimal rendering parameters for a dataset and a target image. Among important rendering parameters, transfer functions are arguably the most effective in highlighting and occluding features in a dataset. They reveal the different faces of the data. In this work, we used opacity transfer functions as the rendering parameter of choice.

The metric module provides a robust distance metric between images using a trained Siamese network described in Section \ref{sec:metric}. The optimizer (Section \ref{sec:optimizer}) uses the distance metric to iteratively assess and refine transfer functions until convergence or for a user-defined number of iterations. Our system computes the optimization process in parallel, discussed in Section \ref{sec:parallel}.

\subsection{Similarity Metric}
\label{sec:metric}
Existing similarity metrics capture low-level features between homogeneous input. To compare datasets with photographs we need resilience to diverse input. This use case cannot be easily addressed by unsupervised image similarity metrics. In our system, we use a deep Siamese neural network that can be trained by the user for different scenarios. In the training phase, the Siamese network takes pairs of images (either similar or dissimilar to one another) as input and updates its weights such that the output is minimized for similar pairs and maximized for dissimilar pairs. In the testing phase, the network takes a pair of images and outputs a value representing the distance between them. 

Siamese networks consist of two identical neural network cores. When comparing two images, each goes through one core and is converted into a feature vector. The L2 norm of the two vectors is then calculated as the distance between the two images. The weights in the Siamese network are trained to decrease the L2 norm for similar images and increase it for dissimilar images. All weights are shared between the two cores of the network. This allows the Siamese network to return the same distance for $A-B$ as for $B-A$ where $A$ and $B$ are two images. 

\begin{figure}
	\centering
	\includegraphics[width=0.5\linewidth]{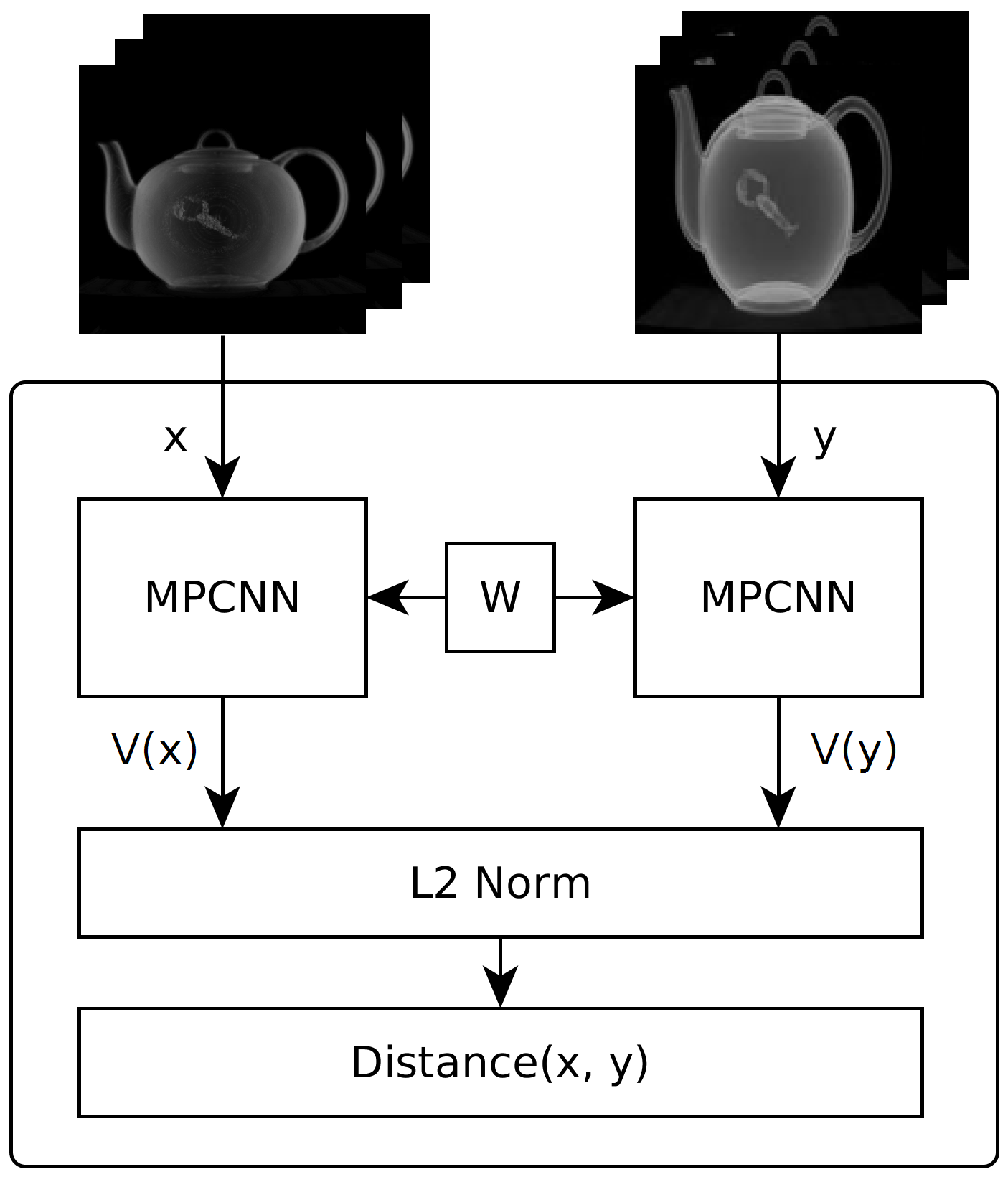}
	\caption{The architecture of our deep Siamese network. Each core of the network is an MPCNN (shown in Figure \ref{fig:cnn}) with shared weights (W). In the training phase, pairs of images enter the network and feature vectors are formed from the cores. 
	}
	\label{fig:arch}
\end{figure}

\begin{figure*}
	\centering
	\includegraphics[width=\linewidth]{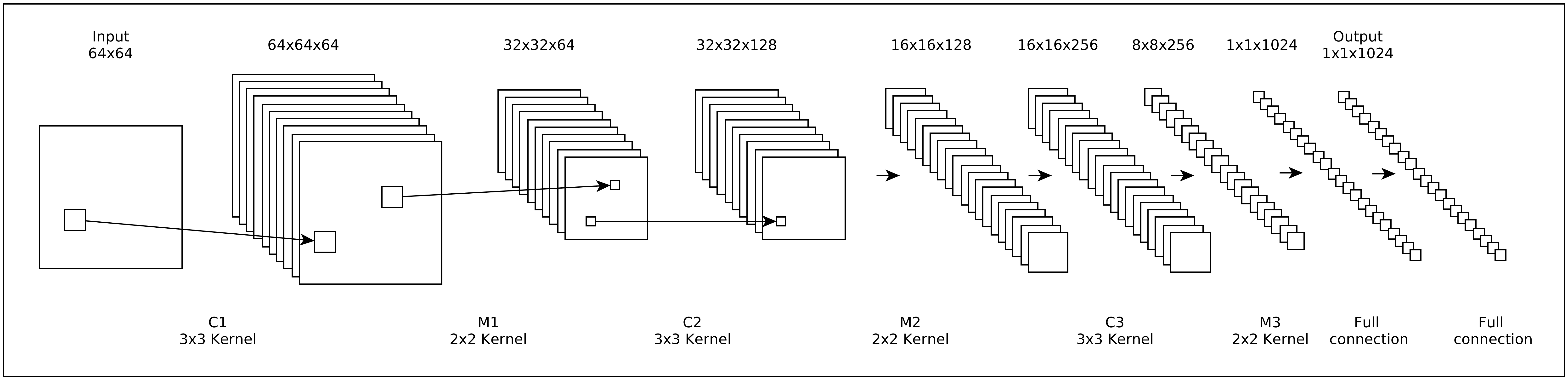}
	\caption{The architecture of the convolutional neural network used in our system. A 64x64 image goes through three convolutional layers (C1, C2, C3), three max-pooling layers (M1, M2, M3) and two fully connected layers to form a feature vector. }
	\label{fig:cnn}
\end{figure*}

Figure \ref{fig:arch} shows the structure of our Siamese network. Each core of our network is a simple max-pooling convolutional neural network (MPCNN) that extracts a feature vector from an input image. Max-pooling is a well known technique that downsamples layers and provides many benefits, such as faster convergence and feature position invariance \cite{ciresan2011flexible}. Many variations of MPCNNs exist. The structure of our MPCNN is shown in Figure \ref{fig:cnn}. The core consists of three convolutional layers and two fully connected layers. Each convolution is followed by max-pooling. The input image, originally of size 64x64, is gradually reduced in size while the convolutional features are increased in depth. In other words, through convolution and downsampling, high level semantics are extracted from low level spatial features. After the final max-pooling layer (M3), 256 8x8 features remain. Similar to other convolutional networks, these features are fully connected to the neurons of the next layer to form a one dimensional feature vector. For each input image, a 1x1x1024 feature vector is then used in the Siamese for L2 norm calculation. 

To train the network, a set of different images is needed. These images can come from different sources. From the set, pairs of similar or dissimilar images are selected by the user for the training phase. We developed a web-based interface to simplify the selection of image pairs (Figure \ref{fig:web}). Initially, the user is presented with a random reference image from the image dataset and a random permutation of those images. The user then selects from the set two images - one highly similar to the reference image and one highly dissimilar. Clicking on \textit{Next} loads a new reference image and the process continues. When done with selecting pairs of images, the user can click on \textit{Submit} to submit the pairs to the system. An Adam optimizer is then used to train the network. 

This step can be time consuming depending on the number of images available. However, training is a one-time step per phenomenon. This implies that a trained Siamese network is potentially reusable by many users who need to work on similar datasets of the same phenomenon.


\subsection{Evolutionary Optimization}
\label{sec:optimizer}

The task of searching for optimal rendering parameters can be mapped to a genetic algorithm (GA) search. Using GA, we derive the optimal transfer function from a random initial set. For every generation of the algorithm, we calculate the fitness of each individual (a transfer function) and pick a set of the best solutions found. We then combine individuals to create better solutions for the next generation. Using a GA allows us to cover a wide range of the search space as well as enabling parallelization inside each iteration of the search process. 

The optimization process starts by using a seeded population of transfer functions and renders the dataset with each of them. The optimizer then evaluates the resulting renders by comparing them to a user-defined target image. Good candidates are then selected from the population. They are randomly mutated and mixed to create a new refined population for the next iteration of the algorithm. 

\begin{algorithm}
    \caption{Evolutionary search for finding the optimal transfer function.}
    \label{alg:search}
    \begin{algorithmic}[1]
        \Procedure{Search}{dataset, target, nGens, popSize}
            \State $population\gets $\textsc{initialPopulation}$(popSize)$
            \For{$g\gets 1, nGens$}
            	\For{$i\gets 1, popSize$}
            		\State $img\gets $\textsc{render}$(dataset, population[i])$
            		\State $fitness\gets $\textsc{evaluate}$(target, img)$
            		\State $population[i].fitness = fitness$
            	\EndFor
            	\State $offsprings\gets $\textsc{select}$(population)$
            	\State $offsprings\gets $\textsc{crossover}$(offsprings)$
				\State $offsprings\gets $\textsc{mutate}$(offsprings)$
				\State $population\gets offsprings$
			\EndFor
			\State \textbf{return} \textsc{selectBest}$(population)$
        \EndProcedure
    \end{algorithmic}
\end{algorithm}

To model our search problem with a GA approach, we first define a discrete version of our transfer function as the constant step function
\begin{equation}
	tf(x) = \sum_{i=0}^{n}{\alpha_{i} \chi_{A_{i}}(x)} 
	\label{eq.tf}
\end{equation}
where $\chi_{A_{i}}$ is the indicator function and $A$ is the set of domain intervals for which the transfer function is defined. In our work, the maximum number of domain intervals is set to 256. This is equal to the number of intervals in the freeform opacity transfer function in the VisIt application \cite{VisIt}. When searching, the number of intervals is constant. We can simply write Equation \ref{eq.tf} as a list of $\alpha_{i}$ values, the \textit{list representation}. This discrete encoding is suitable for the genetic algorithm approach. The overall algorithm is shown in Alg. \ref{alg:search}. The input parameters are the volume data, the target image used in the evaluation function, the number of generations to run the algorithm for and the total size of the population of transfer functions.

In every iteration of the algorithm, the \texttt{RENDER} function renders the dataset using all individuals and a user-defined viewpoint. The resulting renders are then sent to the \texttt{EVALUATE} function where the similarity metric is used to return a cost value for each rendering. The individuals and their costs are then used to create the next generation. New individuals are made from combining the best individuals of the previous population. We chose a ternary tournament selection method to choose the best transfer function among three random individuals of the population and fill the new population with such best individuals. Individuals are then randomly paired, crossed over and mutated in order to create a better generation. For the crossover operation, we used a standard two-point crossover with a probability of 0.8. Mutation was done with a probability of 0.3. To mutate a transfer function, we changed each opacity value to a random integer between 0 and 256 with a probability of 0.05. 

The search space for finding the optimal transfer function is extremely large. Since the range for opacity is typically $[0, 255]$, even with the discrete representation there are $256^{256}$ unique transfer functions. A discrete optimization algorithm cannot easily search in this huge space without uniform coarsening. The need to re-render the dataset for each individual means that choosing a large population is impractical. On the other hand, a small population could mean not covering a large portion of the search space and falling into local minima. To combat these issues we can constrain the transfer function and seed the initial population to control the search space. 

\subsubsection{Controlling the Search Space}
\label{sec:reduction}
We consider transfer functions as a typical 2-dimensional space. The x-axis is the data value. The y-axis is the opacity value. To control the size of the search space and consequently increase search speed, we apply two constraints on our transfer functions. 

First, instead of having 256 discrete values on the x-axis, we coarsen it. For example, the x-axis can be divided into 8 ranges that are 32 wide. Within each range, the y-value (opacity) will be the same. This coarser granularity allows for a smaller population size when performing a search as there are fewer combinations of values to form individuals. We have found that 16 ranges on the x-axis provides a good balance of detail and speed. Initially these ranges are all of equal width. When creating a population of transfer functions, the neighboring ranges will be jittered randomly as sliding windows. When rendering, we smoothen the transfer functions with the smoothing kernel used in VisIt, $0.2 V_{i-1} + 0.6 V_{i} + 0.2 V_{i+1}$, for each $V_i$ value in the transfer function.

Second, we coarsen the y-axis. We initially experimented with 8 equally spaced points: $(0, 32, 64, 96, 128, 160, 192, 224)$ and added 255 to represent full opaqueness. However, after experimenting with the discrete-range transfer functions, we observed that renderings are much more affected by low opaqueness values than by high values. This is mainly because opaqueness accumulates when volume rendering and can quickly reach its maximum \cite{cai2013automatic}. For example, two back-to-back voxels with opacity values of 128 would occlude all voxels behind them. 

With this in mind, we focused on lower opacity and limited the initial range to $\{0, 1, 16, 64, 128\}$. 0 and 1 capture low opacity, 128 captures medium opacity, and 16 and 64 are results of manual tuning. These preset values are for setting initial opacity only. With genetic algorithm, the mutation stage will recombine and crossover individuals. In that stage, fine grained values in the full range of $\{x \in \mathbb{N} \mid 0 \leq x \leq 255\}$ will be recreated. This design choice allows the search to start with a diverse set of coarsely divided range values that gradually become finer throughout the search iterations. 

\subsubsection{Initial Population of Transfer Functions}
\label{sec:initialization}
Depending on the distribution of values in a dataset, many regions in a transfer function may have little to no effect on renderings. Knowing this, we improve the search path using a well-known method in genetic algorithm literature known as population seeding. In seeding GAs, the initial population is not randomly created. Instead problem-specific information is used to create a more promising initial population \cite{julstrom1994seeding}. 

\begin{figure}
	\centering
	\begin{subfigure}{.49\linewidth}
		\includegraphics[width=\linewidth]{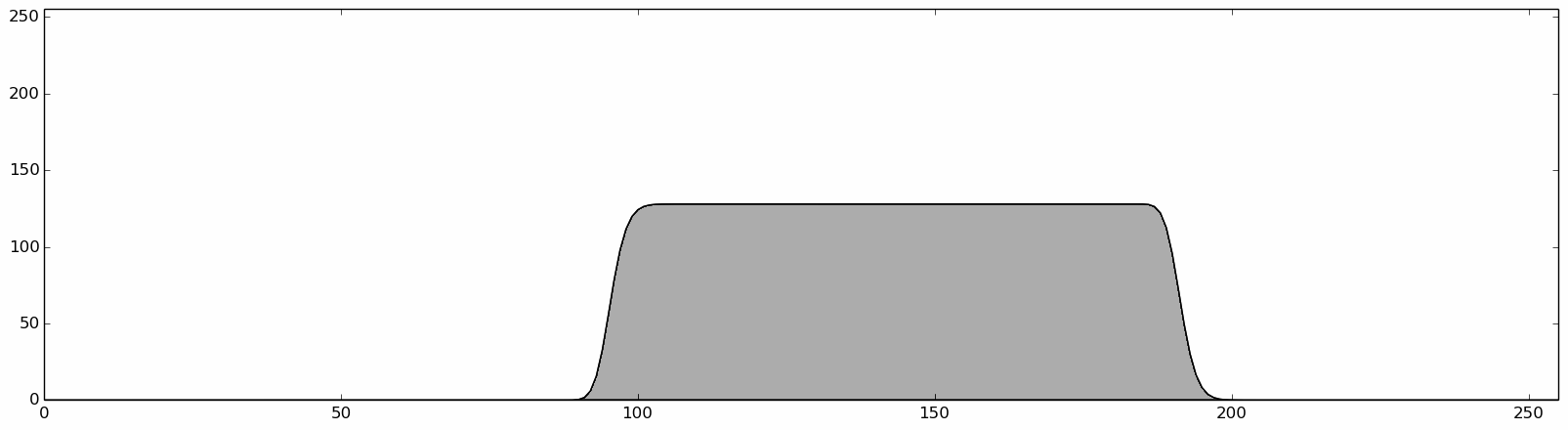}
	\end{subfigure}
	\begin{subfigure}{.49\linewidth}
		\includegraphics[width=\linewidth]{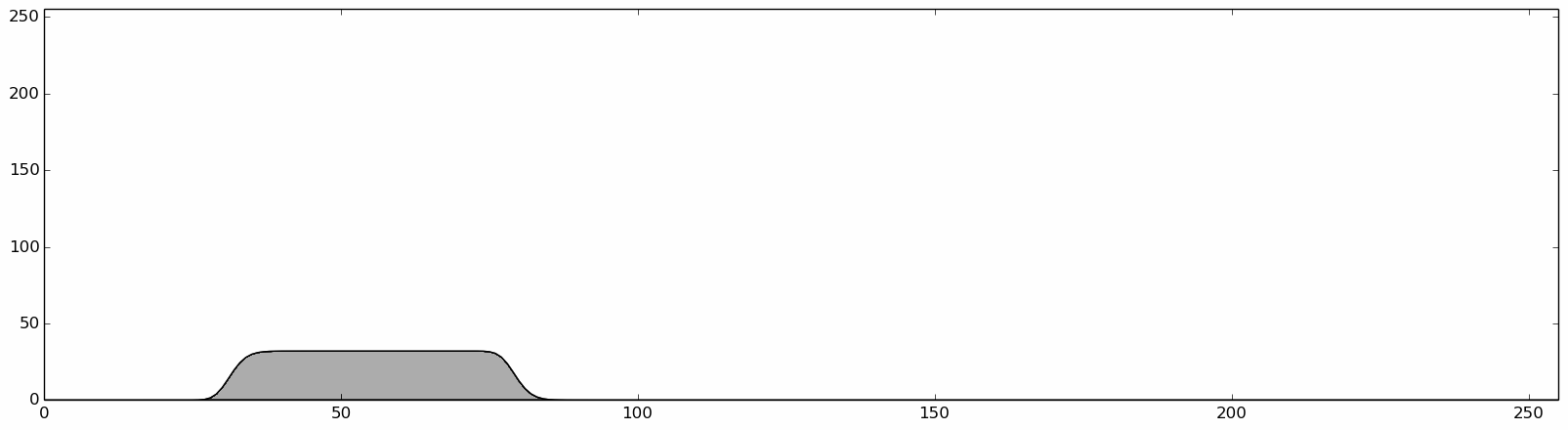}
	\end{subfigure}
	\caption{Two examples of a sliding window transfer function. Both were smoothed with weighted average smoothing (see Section \ref{sec:reduction}).}
	\label{fig:sliding}
\end{figure}

In order to create a uniformly distributed and meaningful population, we used sliding windows of step-wise functions. An example of one such function after smoothing is shown in Figure \ref{fig:sliding}. Sliding windows of different sizes allow us to cover the different features of the data in the initial population. In order to create the initial population, we first generate a set of sliding windows with all possible sizes. The opacity value for each window is chosen from the coarsened y-axis set (see Section \ref{sec:reduction}). Then sliding windows are randomly picked from this set to fill the initial population. A comparison between using population seeding and random initialization is shown in Section \ref{sec:discussion}. 

\subsection{Parallel Implementation}
\label{sec:parallel}
Qualitative search for a transfer function using images can be slow because of rendering time and image comparison. A serial implementation of the same experiment set was attempted, but was impractically slow. To better control and refine the experiment, we parallelized the process. This is especially effective for the genetic algorithm because the evaluation of each generation's transfer functions are completely independent of one another.

\begin{figure}
	\centering
	\includegraphics[width=.7\linewidth]{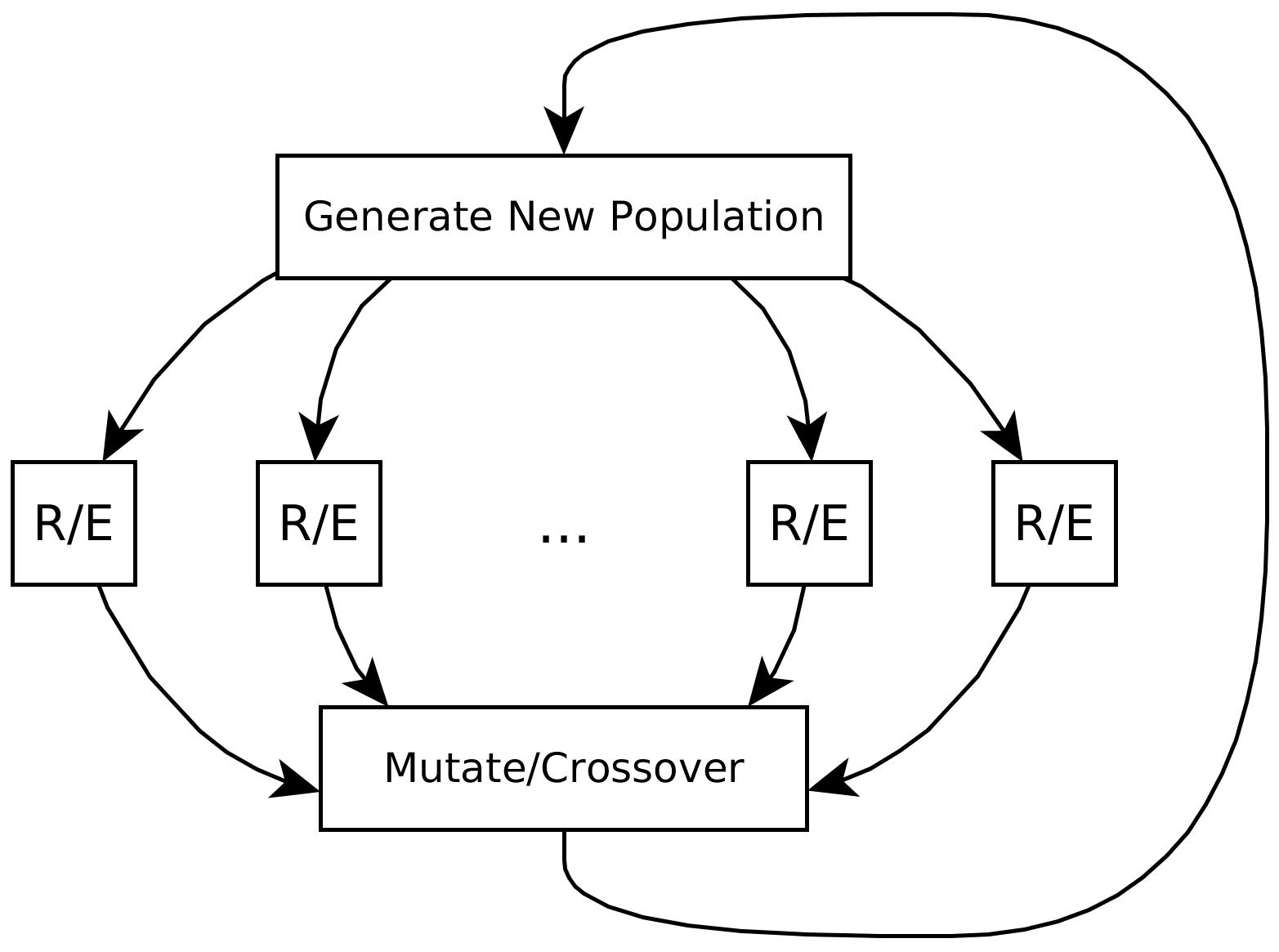}
	\caption{An overview of our parallel implementation. Every R/E node is responsible for rendering the dataset with a set of given transfer functions and evaluating the resulting renderings. The cost of each function is then returned to the master node.}
	\label{fig:parallel}
\end{figure}

Our parallelization is based on MPI (Figure \ref{fig:parallel}) using a cluster of computing nodes. At the beginning of each iteration the population of transfer functions is divided and the members are scattered to a rendering/evaluation (R/E) node where the respective functions from Alg. \ref{alg:search} take place. The results of the evaluation are cost values for each transfer function. These values, accompanied by their respective transfer functions, are then gathered back to the master node where crossover and mutation take place to create the next population. In this design, every R/E node contains a volume rendering process as well as the trained Siamese model that is used for evaluation. The details of our setup are in Section \ref{sec:setup}.

\section{Results}
\label{sec:results}
In this section we demonstrate the capability of our approach on real scientific data using actual photographs of natural phenomena for training. 

\subsection{Environment}
\label{sec:setup}
The two computational components of our approach are the training phase of the neural network and the running of our evolutionary algorithm to generate a transfer function. We used VisIt to load and render the data. We chose 3D texturing as a fast rendering technique. We used VisIt's Python interface to update transfer functions and generate renderings for evaluation. The deep learning component of our work was implemented using the Tensorflow framework's Python API. The VisIt and Tensorflow processes run once in each node and live until the end of the search to avoid startup overhead in each generation. In our implementation the dataset resides in a shared filesystem for all R/E nodes to access.

\begin{figure}
	\centering
	\begin{subfigure}{.49\linewidth}
		\centering
		\includegraphics[width=\linewidth]{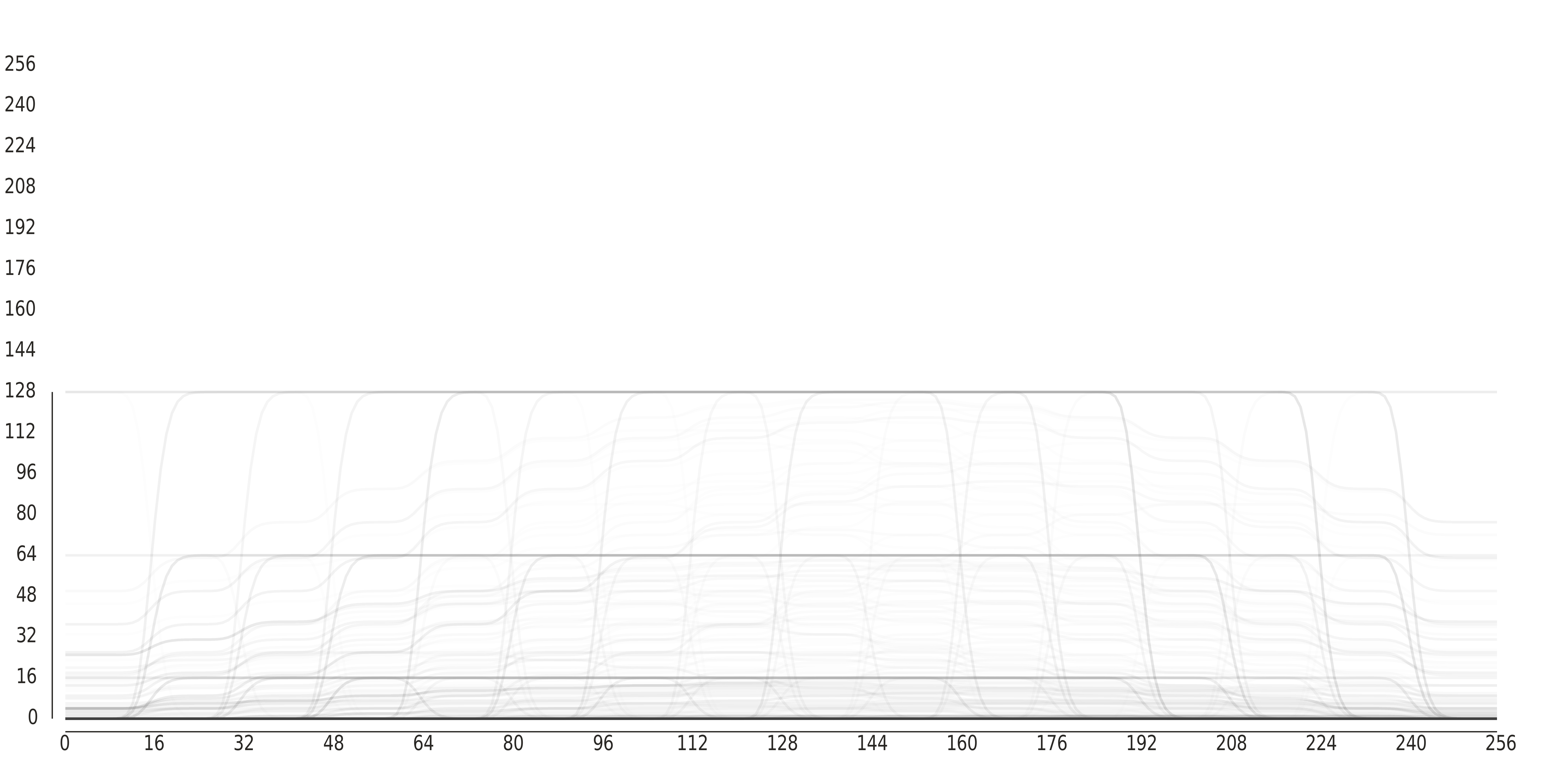}
		\caption{}
		\label{first}
	\end{subfigure}
	\begin{subfigure}{.49\linewidth}
		\centering
		\includegraphics[width=\linewidth]{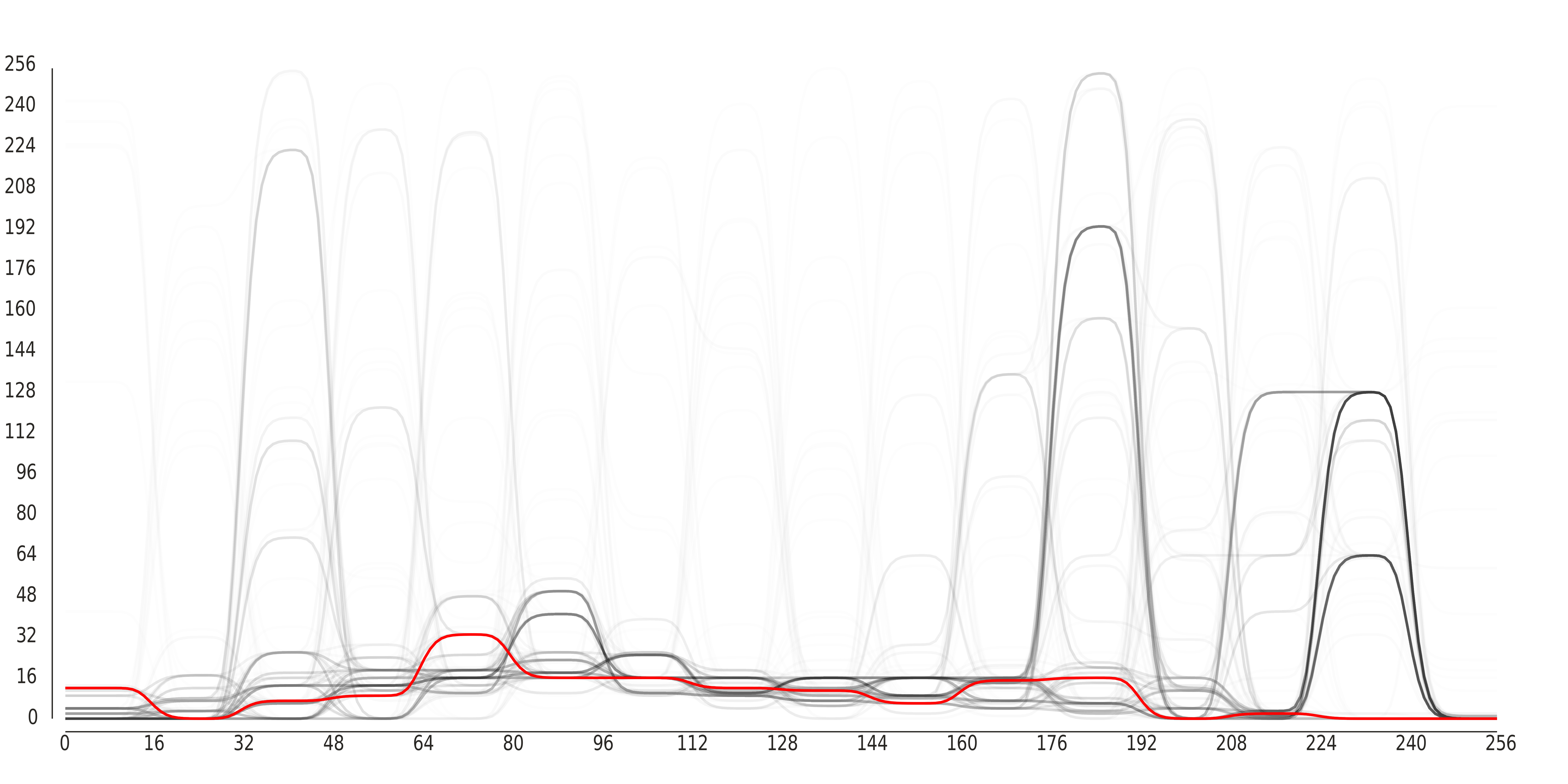}
		\caption{}
		\label{last}
	\end{subfigure}
	 \caption{ (a) Initially seeded initial transfer functions vs. (b) converged transfer functions in the last generation. Note there is still mutation present in the last generation although most transfer functions have converged to the solution (shown in red). }
	 \label{fig:first_last_generations}
\end{figure}

We ran our system for a population size of 600 with 20 generations on 60 nodes of the Blue Waters Supercomputer \cite{bluewaters}. During testing, we ran with a population of 1000, but used 600 for the results due to allocation limitations on Blue Waters. Every node had an AMD 6276 Interlagos processor running at 2.3 GHz and an NVIDIA K20X GPU. 

\subsection{Performance} 
The stochastic computation can be overwhelming. The computation needed to be done in parallel. The calculations in each generation of genetic algorithms are independent of one another and are embarrassingly parallel. The average runtime over 20 tests for the parallel version was 6 minutes and 46 seconds. The main bottleneck was rendering of the volume with each individual. The training phase had a one-time cost of about 30 minutes for choosing pairs of images, and 24 minutes for training the Siamese network for 100 epochs on a single node of Blue Waters. 

\subsection{Highlighting Features in Scientific Data}
We tested our method on real photographs and a scientific dataset. We used a superstorm simulation dataset with a size of $254\times254\times37$ and 48 timesteps. In the results shown in this paper we used timestep 40 and the humidity variable. 

The Siamese network was trained on relevant images, found on Google Images, using search terms such as ``hurricane'' and ``superstorm''. For this dataset we used 89 images for training and derived 342 image pairs using the training interface.
Figure \ref{first} shows a typical seeded population of transfer functions. After 20 generations, the population of transfer functions seems to converge well on a repeatable basis. Figure \ref{last} shows the final population. Note that the final population still contains outlier individuals due to mutation, but there is high convergence towards the solution (shown in red). The final generation contained the globally lowest cost individual which was chosen as the solution. 

\begin{figure}
	\centering
	\begin{subfigure}{0.98\linewidth}
		\frame{\includegraphics[width=\linewidth]{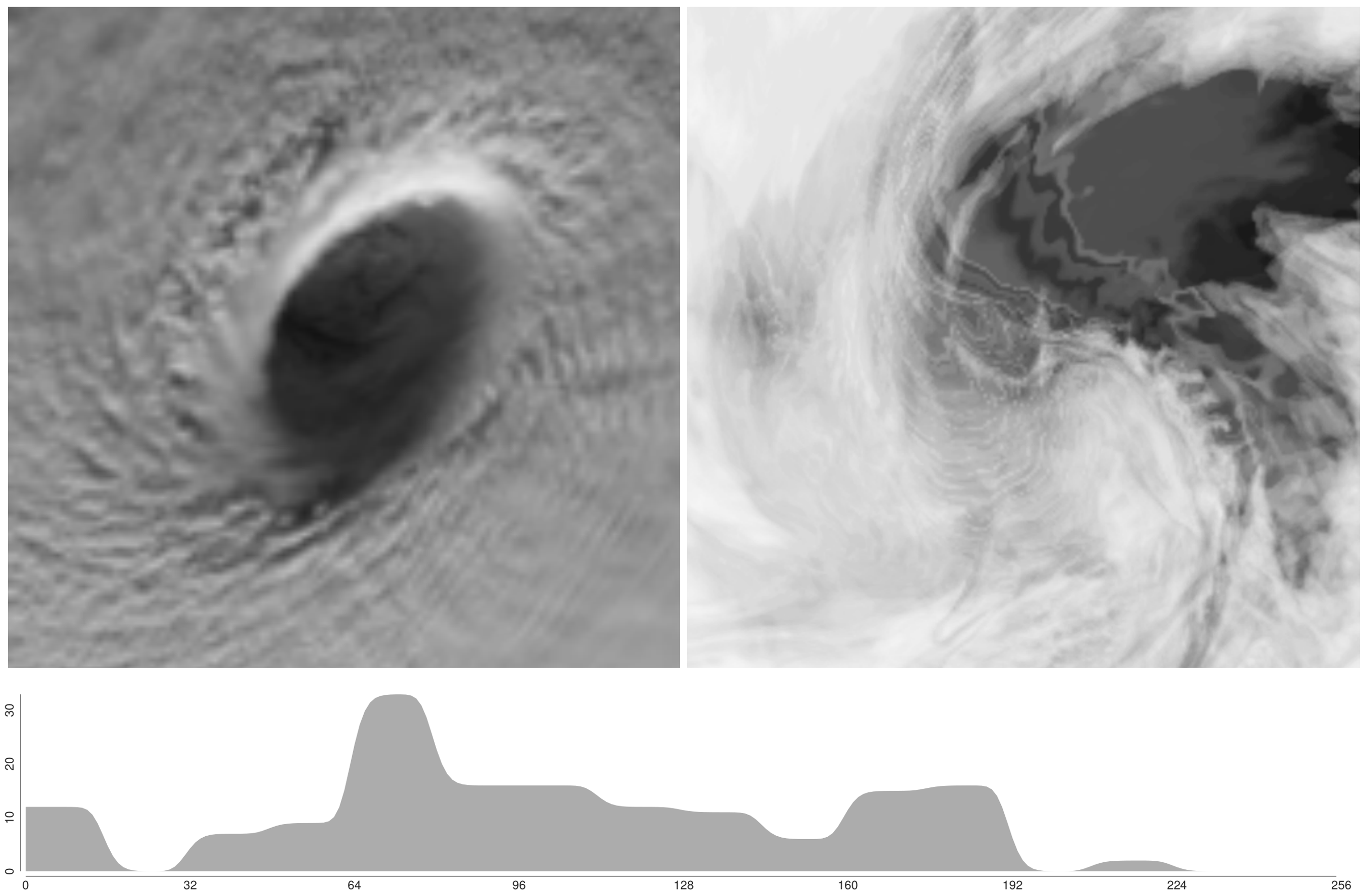}}
		\label{fig:superstorm1}
		\caption{}
	\end{subfigure}
	\begin{subfigure}{0.98\linewidth}
		\frame{\includegraphics[width=\linewidth]{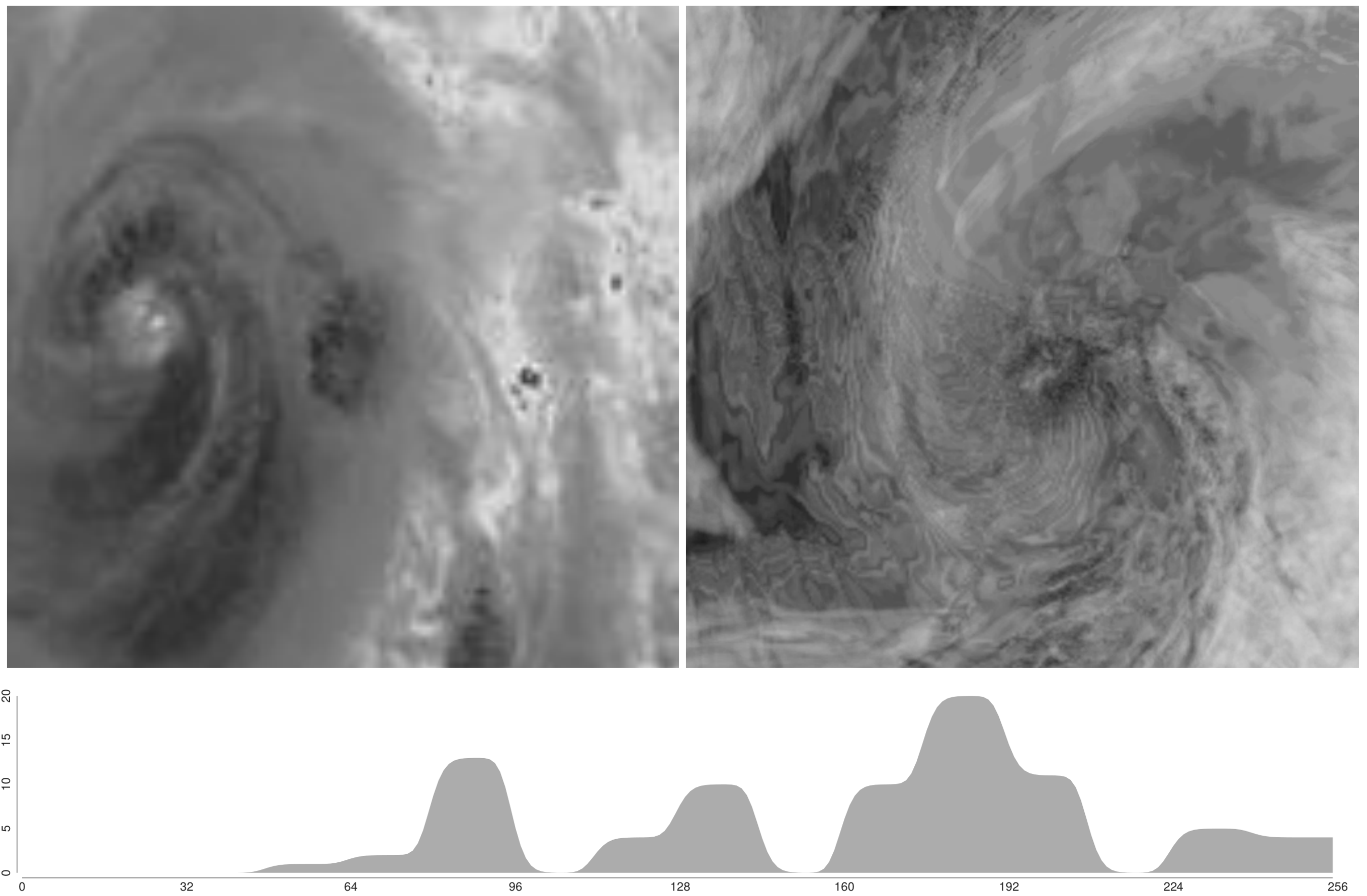}}
		\label{fig:superstorm2}
		\caption{}
	\end{subfigure}
	\begin{subfigure}{0.98\linewidth}
		\frame{\includegraphics[width=\linewidth]{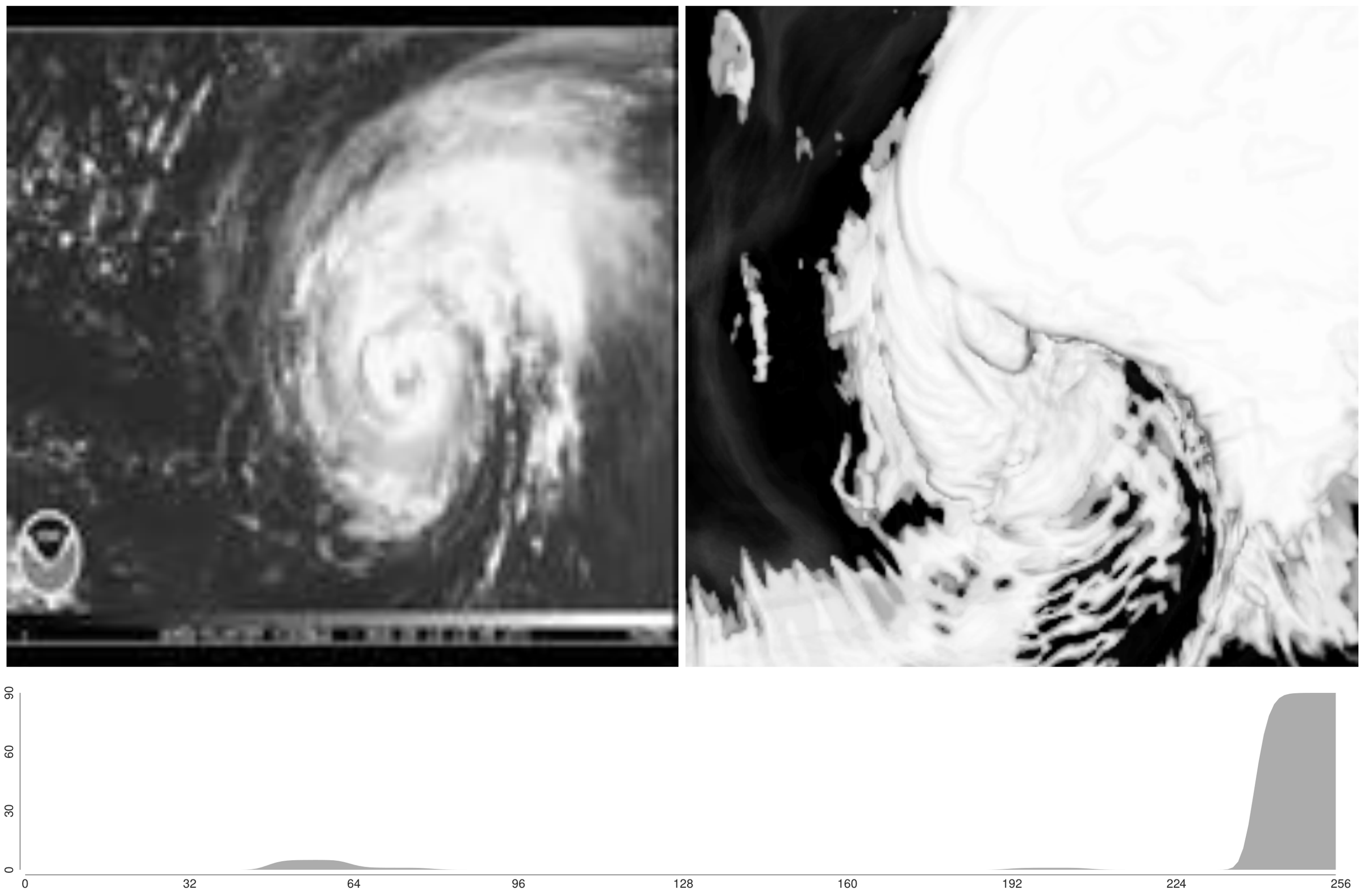}}
		\label{fig:superstorm3}
		\caption{}
	\end{subfigure}
	\caption{ Three real-world target images (left side) along with their respective render of the same volume from the superstorm dataset. These renderings are from the same volume with different transfer functions. The top result was rendered using 3D texturing while the other two were rendered using ray casting. The lowest cost transfer function, used to generate each render, is shown beneath each pair. The distances between the matches were 0.048, 0.026, and 0.080, respectively.} 
	\label{fig:superstorm}
\end{figure}

The results are shown in Figure \ref{fig:superstorm}. The left side shows low-resolution images from the set of online images. The right side shows renders of one volume with three different transfer functions. Each transfer function was generated with the corresponding online image as the target. 

The top result shows a structure resembling the eye of a hurricane. When given the second target image (middle row), the results expose a circular feature close to center of the same storm. The third target image was chosen to be extremely different from the dataset. Despite the difference, we can see that it exposed a boundary to the left of the volume as well as the rain-band structures in the bottom right. These results show that a single volume of the superstorm data exhibits multiple coexisting features similar to three different photographs of natural phenomena. The final distances between the results and their target was 0.048, 0.026, and 0.080, respectively. The result from the second target had the least distance, indicating its high resemblance to the target image. 

%

\section{Discussion and Limitations}
\label{sec:discussion}
\textbf{Overfitting.} With small training data, deep neural networks can be easily overfit. In most circumstances, this reduces the accuracy of the network when presented with new input. Overfit networks are susceptible to small changes in input. However, in our specific use case, we have a lesser need for high recognition accuracy between a pair of input images. The main reason is that there are many variations of input surrounding the correct solution and the genetic algorithm finds and tests these with the metric module. This means that when the metric module leans towards the correct solution, even if slightly, the genetic algorithm can be guided correctly. Additionally, this allows the sensitivity of the overfit DNN be satisfied through small tweaks to the input, done by the GA. The combination between a GA and a DNN helps in this case despite the rough accuracy of the DNN. The reason other similarity metrics don't work despite their inaccuracies, is that they are mostly used for extremely similar images and are not resilient to rotation and movement of features. This is mainly because they do not benefit from convolution whereas DNNs do.

\begin{figure}
	\centering
	\begin{subfigure}{.03\linewidth}
		\centering
		\includegraphics[width=\linewidth]{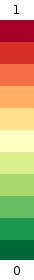}
		\label{colormap}
	\end{subfigure}
	\begin{subfigure}{.28\linewidth}
		\centering
		\includegraphics[width=\linewidth]{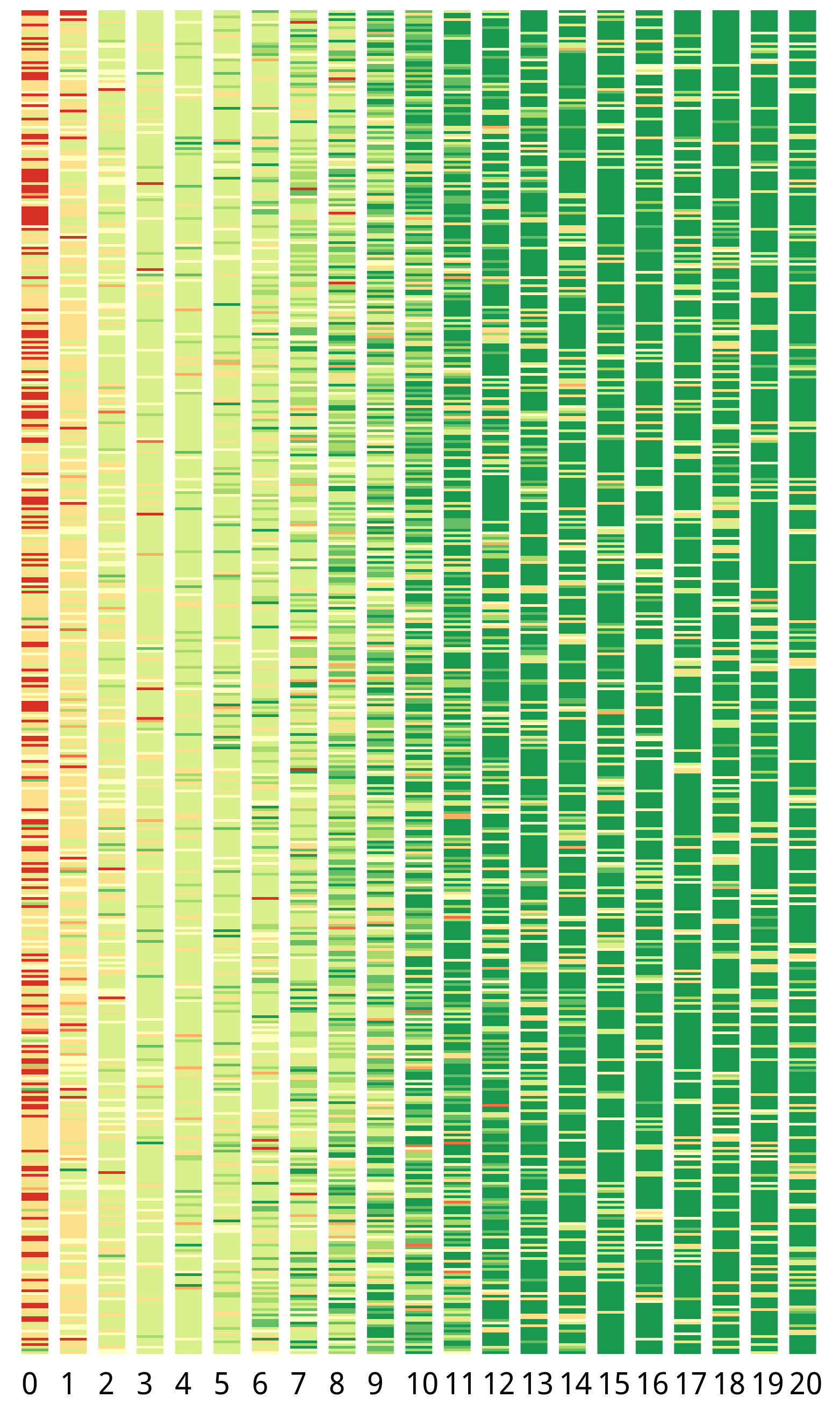}
		\caption{}
		\label{gena}
	\end{subfigure}
	\begin{subfigure}{.28\linewidth}
		\centering
		\includegraphics[width=\linewidth]{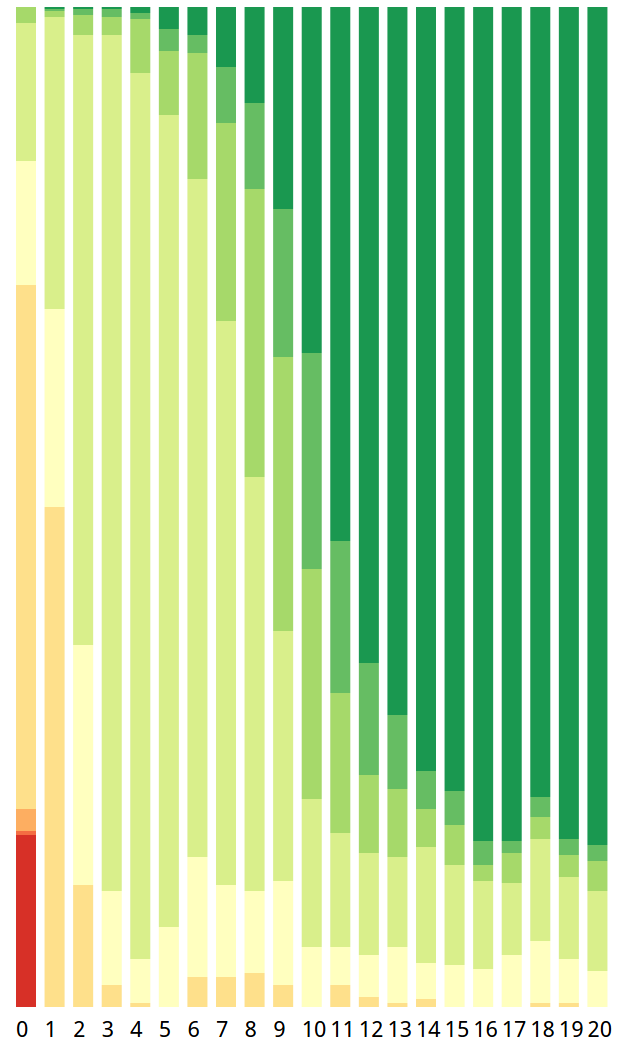}
		\caption{}
		\label{genb}
	\end{subfigure}
	\begin{subfigure}{.28\linewidth}
		\centering
		\includegraphics[width=\linewidth]{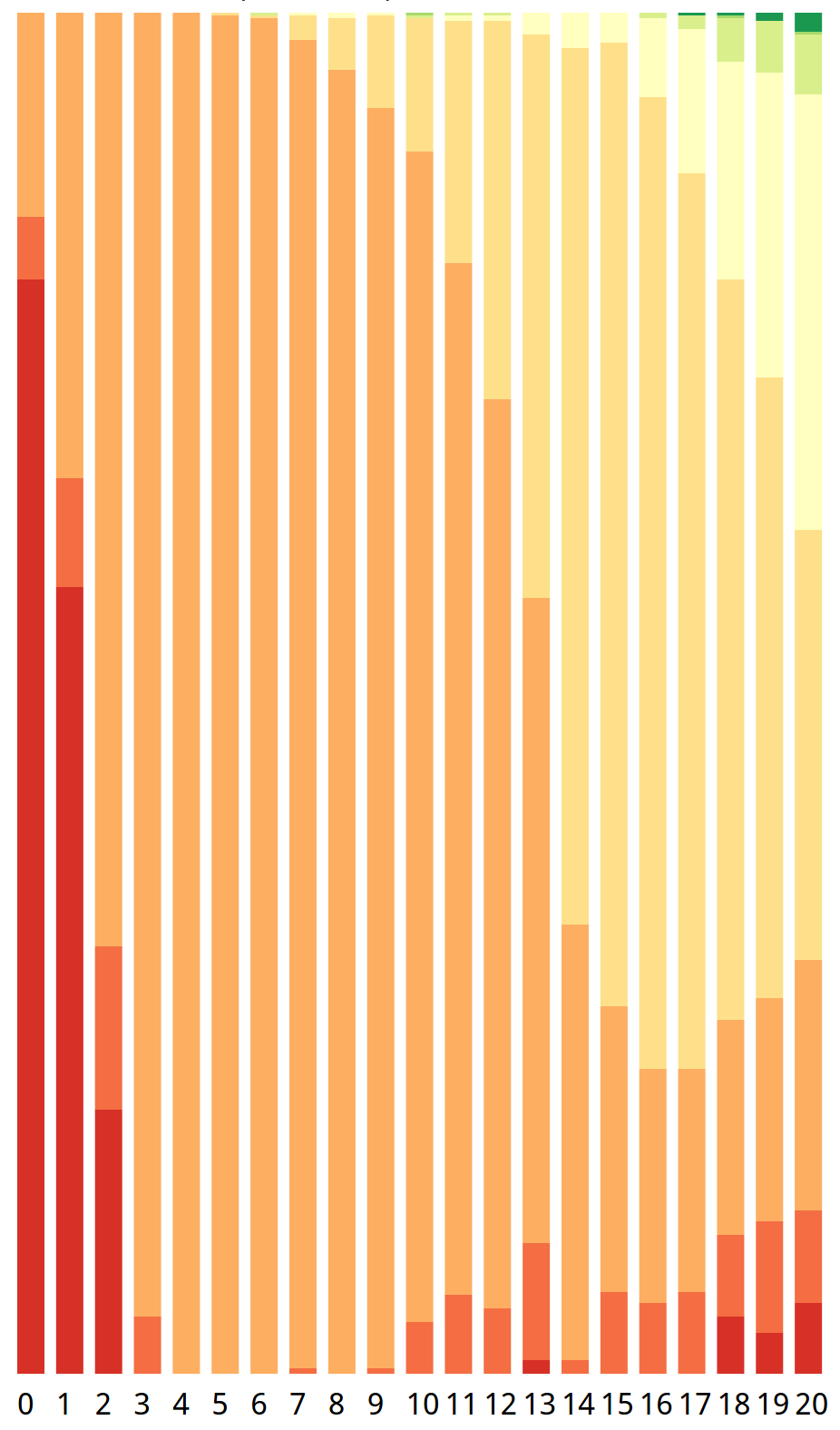}
		\caption{}
		\label{genc}
	\end{subfigure}
	 \caption{Graphs showing the cost of the initial population and the 20 generations of the evolutionary algorithm. Every column in (a) represents a generation and every segment shows the cost of a single transfer function. (b) shows the same generations as (a) but segments in each column are sorted by cost. (c) shows the results of not using population seeding. }
	 \label{fig:generations}
\end{figure}

\textbf{Population seeding.} We ran a sample optimization to test the effects of population seeding, shown in Figure \ref{gena}. Each column represents one generation. Each segment shows the cost of a single transfer function by color. The costs are normalized based on the minimum and maximum costs of the optimization. The first initial population contains a variety of cost values. From generation 10 onward the solution starts to converge to the target. Figure \ref{genb} shows the same graph with each generation sorted based on the cost of its members. The lowest cost was $0.012$. To see the effects of population seeding, we ran the same optimization without seeding the initial population. The overview of the optimization is shown in Figure \ref{genc}. We can see that in 20 generations the optimizer spent most of its time visiting bad solutions. The number of bad solutions also increase from generation 14. The lowest cost was $0.02$. 

\textbf{Data alteration.} Note that although our system searches for a rendered image as similar to the target as possible, it does not change the data itself. The only parameter that changes is the transfer function. Our system aims at automating a customization of the transfer function similar to how a human would, while the data itself is kept intact. Additionally, similar to a user's search and other GA-based approaches, finding a global optima cannot be guaranteed. 


\textbf{Camera viewpoint.} One of the parameters that affects our results is the camera viewpoint. However, we argue that this is a constraint of transfer functions. For a single volume, different viewpoints can have different occluded regions, not addressable by a single global transfer function. 

\begin{figure}
	\centering
	\includegraphics[width=.3\linewidth]{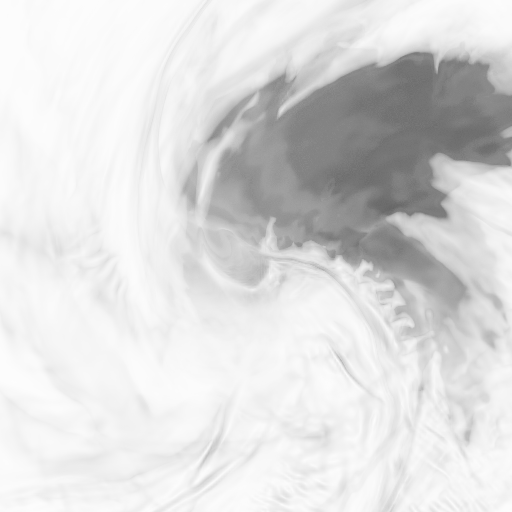}
	\caption{The ray casted version of the superstorm eye result. This image shows that when the rendering method differs from the one used in the optimization, the result may look different. }
	\label{fig:raycasting}
\end{figure}

\textbf{Rendering method.} While opacity transfer functions affect the rendered results, the particular renderer of choice affects the results as well. For maximum consistency, the renderer used in the optimizer should be the same as the render used for the final renderings. In some cases ray casting produced a similar image to 3D texturing, but in some cases it did not (Figure \ref{fig:raycasting}). 

\section{Conclusion and Future Works}
In this work, we have shown that a qualitative input that roughly describes what the user has in mind can automatically drive an optimization to a suitable render. Our system uses a deep neural network's understanding of the phenomena being visualized and a target image to assess the quality of rendering parameters in the optimization. This allows users to search for the existence of a feature in a volume dataset using a target image and delegate the iterative search process to the system. 

We believe the question posed by this work enables a novel visualization-driven capability for searching through datasets (visual search). Additionally, our approach facilitates rendering reproducibility and automatic parameter selection.

In future works, we would like to compare our method with an expert's manual workflow. Also, we would like to explore more specific use cases of our technique. Two potential cases are searching for abnormalities such as tumors in medical datasets, and rendering volume illustrations from an image corpus.

Another direction of future work is to expand our technique to other rendering parameters so that we can reliably automate the rendering pipeline with user knowledge as a backbone instead of an active element. Additionally, removing the training phase by automatically creating the similar/dissimilar pairs for the neural network could be valuable. One potential method could be using unsupervised and discriminative image similarity measures that can only point out highly similar/dissimilar images.

\section*{Acknowledgements}
This research was funded in part by the Blue Waters sustained-petascale computing project, which is supported by the National Science Foundation (awards OCI-0725070 and ACI-1238993) and the state of Illinois. The authors are also supported in part by NSF Award CNS-1629890, USDI-NPS P14AC01485, Intel Parallel Computing Center (IPCC) at the Joint Institute of Computational Science of University of Tennessee, and the Engineering Research Center Program of the National Science Foundation and the Department of Energy under NSF Award Number EEC-1041877. The authors would like to thank the anonymous reviewers for their valuable comments and suggestions. The NVIDIA hardware donation program provided GPU hardware for this project. Mississippi State University provided the WRF simulation data. 

\bibliographystyle{eg-alpha-doi}
\bibliography{main}

\end{document}